\documentclass[sigplan,screen, 10pt, authorversion]{acmart}
\usepackage{cleveref}
\usepackage{cite}
\usepackage{amsmath}
\usepackage{lipsum}  
\usepackage{stmaryrd}
\usepackage{stackengine}
\usepackage{algorithm}
\usepackage{algorithmicx}
\usepackage{algpseudocode}
\usepackage{paralist}
\usepackage{natbib}
\usepackage{xcolor}
\usepackage{xspace}
\usepackage{listings}
\usepackage{latexsym}
\usepackage{fontawesome}
\usepackage{fixfoot}
\usepackage{tikz}
\usepackage{float}
\usepackage{multicol}
\usepackage{booktabs}
\usepackage{caption}
\usepackage{array,ragged2e}
\usepackage{varwidth,setspace}
\usetikzlibrary{arrows}
\usetikzlibrary{positioning,decorations.pathmorphing}
\usetikzlibrary{arrows.meta}
\usepackage{subcaption}
\usepackage{hyperref}
\newif\ifdraft
\draftfalse

\newcommand{\scode}[1]{\mbox{\small \texttt{#1}}}

\newcommand{\kcode}[1]{\mbox{\lstinline[language=Kotlin,basicstyle={\small\ttfamily}]^#1^}}

\ifdraft

\fi

\newcommand{\para}[1]{\subparagraph*{\textbf{#1.}}}

\def\signed #1{{\leavevmode\unskip\nobreak\hfil\penalty50\hskip2em
  \hbox{}\nobreak\hfil---#1%
  \parfillskip=0pt \finalhyphendemerits=0 \endgraf}}

\newsavebox\mybox
\newenvironment{aquote}[1]
  {\savebox\mybox{#1}\begin{quote}}
  {\signed{\usebox\mybox}\end{quote}}

\newcolumntype{M}{>{\begin{varwidth}{3cm}}l<{\end{varwidth}}} %

\AtBeginDocument{%
  }
    \setcopyright{acmlicensed}
\copyrightyear{2024}
\acmYear{2024}
\acmDOI{XXXXXXX.XXXXXXX}

\definecolor{orange}{rgb}{0.99,0.48,0.13}
\definecolor{hilaorange}{rgb}{0.81,0.39,0.10}
\definecolor{grass}{rgb}{0.18,0.80,0.18}
\definecolor{ForestGreen}{rgb}{0.133, 0.545, 0.133}

\definecolor{codegreen}{rgb}{0,0.35,0.1}
\definecolor{codegray}{rgb}{0.5,0.5,0.5}
\definecolor{codepurple}{rgb}{0.58,0,0.82}
\definecolor{backcolour}{rgb}{0.95,0.95,0.92}
\definecolor{cobalt}{rgb}{0.0, 0.28, 0.67}
\definecolor{numgreen}{RGB}{19,137,103}
\definecolor{stringred}{RGB}{163,21,21}

\lstdefinelanguage{Kotlin}{
  comment=[l]{//},
  emph={filter, firstOrNull, forEach, lazy, map, mapNotNull, println},
  identifierstyle=\color{black},
  keywords={!in, !is, abstract, actual, annotation, as, as?, break, by, catch, class, companion, const, constructor, continue, crossinline, data, delegate, do, dynamic, else, enum, expect, external, false, field, file, final, finally, for, fun, get, if, import, in, infix, init, inline, inner, interface, internal, is, lateinit, noinline, null, object, open, operator, out, override, package, param, private, property, protected, public, receiveris, reified, return, return@, sealed, set, setparam, super, suspend, tailrec, this, throw, true, try, typealias, typeof, val, var, vararg, when, where, while},
  morecomment=[s]{/*}{*/},
  morestring=[b]",
  morestring=[s]{"""*}{*"""},
  ndkeywords={@Deprecated, @JvmField, @JvmName, @JvmOverloads, @JvmStatic, @JvmSynthetic, Array, Byte, Double, Float, Int, Integer, Iterable, Long, Runnable, Short, String, Any, Unit, Nothing},
  ndkeywordstyle={\color{ForestGreen}\bfseries},
  sensitive=true,
  moredelim=**[is][{\color{black!85}}]{^}{^},
}

\lstdefinestyle{sharpc}{language=[Sharp]C}

\lstdefinestyle{mystyle}{
    language=Kotlin,
    commentstyle=\color{codegreen},
    keywordstyle=\bfseries\color{cobalt},
    numberstyle=\tiny\color{codegray},
    stringstyle=\color{stringred},
    basicstyle=\ttfamily\footnotesize,
    breakatwhitespace=false,         
    breaklines=true,                 
    captionpos=b,                    
    keepspaces=true,                 
    numbers=left,                    
    numbersep=5pt,                  
    showspaces=false,                
    showstringspaces=false,
    showtabs=false,                  
    tabsize=2,
    mathescape
}
\begin{document}
\title{Kotlin's Type System is (Also) Unsound}

\author{Elad Kinsbruner}
\email{kinsbruner@cs.technion.ac.il}
\affiliation{%
  \institution{Technion}
  \city{Haifa}
  \country{Israel}
}

\author{Hila Peleg}
\email{hilap@cs.technion.ac.il}
\affiliation{%
  \institution{Technion}
  \city{Haifa}
  \country{Israel}
}

\author{Shachar Itzhaky}
\email{shachari@cs.technion.ac.il}
\affiliation{%
  \institution{Technion}
  \city{Haifa}
  \country{Israel}
}

\renewcommand{\shortauthors}{Kinsbruner et al.}



\begin{abstract}
Soundness of a type system
is a fundemental property
that guarantees that no operation
that is not supported by a value
will be performed on that value at run time.
A type checker for a sound
type system is expected to
issue a warning on every
type error.
While soundness is a desirable
property for many practical
applications,
in 2016, Amin and Tate presented
the first unsoundness proof
for two major industry languages: Java
and Scala.
This proof relied on
use-site variance and implicit null values.

We present an unsoundness proof
for Kotlin, another emerging industry language,
which relies on a previously unknown
unsound combination of language features.
Kotlin does not have implicit null values,
meaning that the proof by Amin and Tate would not
work for Kotlin.
Our new proof, which is an infringing
code snippet, utilizes Kotlin's \emph{declaration-site}
variance specification and does not
require implicit null values.

We present this counterexample to 
soundness in full
along with detailed explanations
of every step.
Finally, we present a thorough
discussion on precisely which language
features cause this issue,
as well as how Kotlin's compiler
can be patched to fix it.
\end{abstract}

\maketitle
\section{Introduction}
Type systems are a common abstraction used
by most modern programming languages,
and ensuring their soundness is a major design goal. %
A type system is said to be \emph{sound}
if whenever a program is well typed,
it will never perform on a value an operation 
that is not supported by that value.
This is a basic yet useful guarantee
to have for programs. 
For real world languages, a compiler
for a language with a sound type system
will issue a warning or an error for programs
that are not well-typed.
Some theoretical programming languages,
like the Simply-Typed Lambda Calculus
have been proven sound~\citep{pierce2002types},
as were some more widespread
languages such as the standard dialect of ML~\citep{vaninwegen1997towards}.
Soundness results have practical importance
as unsound type systems may cause unexpected
runtime errors.
In 2016, Amin and Tate proved~\citep{10.1145/2983990.2984004}
a major unsoundness result for two popular industry languages
--- Java and Scala.

Java's unsoundness proof relies on Java's use-site
variance designation~\citep{javavariancespec}, 
which means that generic types can have a lower or upper
bound instead of a precise type.
The proof crucially also relies on 
implicit null pointers
that serve as members of every class type,
including types that otherwise would have been uninhabited.
Implicit null pointers were solely blamed by the authors,
including in the paper's subtitle: ``The
Existential Crisis of Null Pointers''.
The paper also refers to implicit null pointers
as the
``billion-dollar mistake'',
paraphrasing a talk~\citep{billion09} by Hoare:

\begin{aquote}{Nada Amin \& Ross Tate~\citep{10.1145/2983990.2984004}}
[Java's] reasoning for wildcards and path-dependent types 
would be perfectly valid if not for implicit null values.
[...] recent industry languages such as Ceylon, Kotlin, 
and Rust have already made nulls explicit in 
reaction to the problems that implicit nulls have caused [...].
\end{aquote}

The Kotlin programming language~\citep{kotlinlang} is a modern
industry language developed by JetBrains.
Kotlin's popularity increased in recent years
according to developer surveys~\citep{sosurvey},
and 9.06\% of surveyed developers used the language in 2023
(up from 7.8\% in 2020).
Kotlin, like Java and Scala, is a JVM language and one of its declared
design aims is to fix some of Java's shortcomings~\citep{kotlinspec},
including implicit nulls.
Kotlin's type system, designed by Tate~\citep{tate} and Andrey Breslav,
requires types to be explicitly denoted as either nullable
or non-nullable, and enforces this requirement.
Thanks to this, 
Kotlin's type system does not
suffer from the pitfall that caused Java's unsoundness.
Nonetheless, we will see that Kotlin's type system is also
unsound, for reasons not related to implicit nulls.
This result stems from Kotlin features that are
not present in Java, so our example cannot be translated to Java.

In this paper, we will show how Kotlin's \emph{declaration}-site
variance allows us to treat
a non-variant generic class as a variant one.
Abusing variance is also at the heart of Java's unsoundness
proof. 
However, Kotlin's declaration-site variance ``buries''
variance information that should make the coercion illegal,
making it harder for the compiler, the IDE and the user
to notice that anything went wrong.
Indeed, both Kotlin's compiler
and the latest version of JetBrains' IntelliJ IDEA, 
the official IDE for Kotlin~\citep{intellij}, don't even issue
a warning on the offending code.
Even worse, compiler implementation details cause the runtime error 
to not occur when performing
the coercion, but rather only when attempting to perform
an illegal operation on the coerced object.

This result has implications on existing Kotlin code.
\Cref{sec:variations} will discuss how this issue can be subverted as
a programmer looking to write type-safe code. 
In addition, \Cref{sec:variations} will discuss which language 
features affect this issue and present small improvements
to the compiler will allow it to recognize
this kind of potentially unsafe code and warn the programmer.

Kotlin's unsoundness in general is not
a novel result; 
another cause for unsoundness in Kotlin (also variance related)
has previously been documented in Kotlin's bug tracker~\citep{ischeck}.
Our counterexample \emph{does not}
rely on the same features as the issue
from the Kotlin bug tracker.
Fixing that issue would not be a major breaking change
to the language, 
and issuing a warning would be even simpler:
simply check if a (smart) cast
is attempted between a variant parent class
to a non-variant child class, and issue a warning
if so.
Our counterexample, however,
is fundemental: it cannot be solved without 
making major changes to the language, as
discussed in \Cref{subsec:blame}.
Also, it can't be found easily by a linter,
because its components may be scattered
across different functions.
This matter is discussed in more detail in \Cref{subsec:known}.

In informal conversations with members of
the Kotlin team at JetBrains we confirmed
that this issue was real and should be addressed,
and we will file a bug report for it.

\para{Contributions}
The contributions of this paper are:
\begin{compactitem}
\item A previously unknown unsound combination of language features,
and a new counterexample to soundness proof utilizing these language features.
\item Detailed analysis of the result's causes and implications,
as well as comparison to other popular industry languages.
\item Discussion of how the issue could have been averted from a 
language design perspective, as well as how users can prevent it
in their own code.
\item A suggested fix to the Kotlin compiler to help it
recognize this issue.
\end{compactitem}

\section{Preliminaries}\label{sec:preliminaries}
This section will survey the basic background
required for the technical details in \Cref{sec:technical}
in the interest of self-containment.

\subsection{The JVM Stack Machine}
The \emph{Java Virtual Machine} (\emph{JVM})
is a virtual machine used to run
code compiled to its bytecode, 
\emph{Java bytecode}.
The first language to be compiled
to the JVM is the Java programming language~\citep{javaspec},
from which its name derives.
In the 30 years since the JVM's
introduction, and thanks to
Java's immense popularity,
other languages
that compile to Java bytecode emerged.
These languages, e.g., Scala~\citep{scalalang},
Clojure~\citep{clojurelang} and indeed Kotlin,
enjoy interoperability with Java code
and access to the vast Java ecosystem.
However, these languages necessarily inherit
some of Java's limitations.

Being an abstract machine
that does not have to
run on actual hardware, the JVM
allows for a higher level of
abstraction in programs.
An important example is RTTI:
Run-time Type Information,
which is additional metadata
that the JVM appends to values
so that their dynamic type
can be queried at run time.
This, too, is transparent to 
JVM bytecode, and is handled
directly by the JVM.

JVM objects are tagged with a ``conservative type'',
which may be less specific than the dynamic
type in the object's RTTI.
Conversions to a more specific type
are done with the \scode{CHECKCAST classname}
instruction, which checks the RTTI of its operand, 
and if it can be cast to \scode{classname},
performs the cast.
Otherwise, it throws a \scode{ClassCastException}.
Importantly, this instruction is \emph{not}
only placed by the JVM for explicit casts,
but in any situation in which
the compiler is trying to coerce a value
of a certain type to be a value of
another type.

\para{Type erasure}
The JVM implements generics using \emph{type erasure}.
This means that type parameters are not kept
as part of the RTTI and do not exist
at all at run-time.
Type parameters are verified by the compiler
and then erased, 
which means that, for example, a \scode{List<Int>}
is treated at run time like a \scode{List<*>}.
The compiler is aware of the
type parameters, and the objects in the list
have RTTI witnessing their type, 
and so the compiler can insert \scode{CHECKCAST Int} instructions
after every \scode{List\#get} operation
to cast returned \scode{Any}
objects back to the intended type.
However, since the list object itself
does not have RTTI witnessing it is 
specifically a \scode{List<Int>},
trying to check if an
object is specifically a \scode{List<Int>}
at run-time will ignore the type
parameter and return \scode{true}
if the object is any \scode{List} instance,
including, for example, a \scode{List<String>}.

\subsection{Variance}
Given two classes, \scode{Derived} and \scode{Base},
such that \scode{Derived} is a subclass of \scode{Base},
generic classes using them can have one of three
relations:
\begin{compactitem}
\item \scode{G<Derived>} is a subclass of \scode{G<Base>} (\emph{covariance}),
\item \scode{G<Base>} is a subclass of \scode{G<Derived>} (\emph{contravariance}), or
\item \scode{G<Derived>} and \scode{G<Base>} aren't related (\emph{no-variance}).
\end{compactitem}

For example, consider a \scode{Producer<T>} class
which only has one method that accepts no parameters
and returns an instance of \scode{T}.
Since \scode{Derived} is a subclass of \scode{Base},
returning a \scode{Derived} object in a \scode{Producer<Base>}
shouldn't cause any problem. 
Therefore,
this class should be safely covariant.
Analogously, a \scode{Consumer<T>} class
which accepts a parameter of type \scode{T}
and returns nothing should be safely contravariant.
A class containing both ``consumer-like''
methods and ``producer-like'' methods should be non-variant.

\para{Use- vs.\@ declaration-site variance}
There are two common approaches to specifying
which variance a value exhibits.
The first option, used by Java
and Scala, is \emph{use-site variance specification},
where every class is non-variant by default,
but the type of a specific variable or parameter
can be marked covariant or contravariant
at use site, using syntax like \scode{List<? extends T> param}
in Java or \scode{param: List[+T]} in Scala.
The type checker checks that these
values are used according to their denoted variance.
The other option, used by Kotlin,
is \emph{declaration-site variance specification},
where every class declaration can be marked with its own
default variance, using syntax like \scode{class List<out T>}.
The type checker checks that the class methods
use the type parameter according to its denoted variance.
This option has the advantage of being more convenient:
``producer-like'' classes can always be covariant
and ``consumer-like'' classes can always be contravariant.

In Kotlin, inheriting classes don't have to respect
their parent class's variance specification.
This fact is used in Kotlin's collection hierarchy,
which includes a \scode{List<out T>} interface
with only ``producer-like`` methods such as \scode{get(idx: Int): T},
and a \scode{MutableList<T> : List<T>} interface
extending it with ``consumer-like'' methods such as \scode{add(elem: T): Boolean}.

\noindent Since \scode{MutableList} implementors have both ``consumer''
and ``producer'' methods, it must be non-variant,
despite descending from the covariant interface \scode{List}.

\section{Counterexample to Soundness}\label{sec:technical}
\begin{figure}[t]
\begin{lstlisting}[xleftmargin=1.5em, escapechar=@]
open class B@\label{line:classes-start}@

class A private constructor() : B() {
    fun secretMethod() { /* ... */ }
} @\label{line:classes-end}@

// ...

fun getA(): A {
    val list = mutableListOf<A>()
    val upcast: List<A> = list
    val covariance: List<B> = upcast
    val downcast: MutableList<B> = 
                     covariance as MutableList@\label{line:downcast}@
    downcast.add(B())@\label{line:mutate}@
    return list[0]
}
getA().secretMethod()@\label{line:usage}@
\end{lstlisting}
\caption{A code snippet that demonstrates Kotlin's type system unsoundness}
\Description{A code snippet that demonstrates Kotlin's type system unsoundness}
\label{fig:mainexample}
\end{figure}

After reviewing the relevant features of Kotlin,
we are ready to construct a method \scode{getA}
that appears statically to return an 
instance of the uninhabited type \scode{A},
or, equivalently, appears to coerce an 
instance of type \scode{B}
to be of type \scode{A}. 
The method \scode{getA} is shown in \Cref{fig:mainexample}.
This code type-checks according to the 
Kotlin Language Specification~\citep{kotlinspec} 
and can be compiled by \texttt{kotlinc} version 1.9.23 without
any warnings.

This proof causes an illegal coercion:
from a superclass to a proper subclass.
We create an inheritance hierarchy of two classes \scode{A <: B} 
(lines \ref{line:classes-start}--\ref{line:classes-end}).
We set \scode{A}'s constructor's visibility \scode{private},
which makes \scode{A} uninhabited (this is not necessary
but it distills the heart of the issue).
Since \scode{A} is uninhabited, 
if we find a total function that takes no arguments
and returns \scode{A}, without the use
of reflection or similar tools,
the type system must be unsound. 
Note that this is equivalent to running
an illegal operation on some object,
because conceptually, if the type is 
uninhabited, and we return an object
of that type, it must be some other, 
illegally coerced, object.

\para{Upcast}
\begin{quote}
\begin{lstlisting}[numbers=none]
val list = mutableListOf<A>()
val upcast: List<A> = list\end{lstlisting}
\end{quote}
The function \scode{getA} starts by
initializing a \scode{MutableList<A>}
object. 
As mentioned in \Cref{sec:preliminaries},
\scode{MutableList} inherits from \scode{List}.
This means that upcasting \scode{MutableList<A>}
to \scode{List<A>} must be legal.
Indeed, the compiler does not issue
a warning or an error, and this line
runs successfully at runtime.

Having the standard library \scode{MutableList} 
inherit from \scode{List} is sometimes considered a dubious 
language design decision because \scode{List}
objects may be erroneously considered to be immutable~\citep{twomutables}.
Due to this implicit assumption of immutability,
the \scode{List} interface is marked as covariant.
This leads to catastrophic results:
Since \scode{List} is covariant in its type parameter
but \scode{MutableList} is non-variant, 
this is equivalent to Java code allowing
an upcast from \scode{MutableList<T>}
to \scode{List<? extends T>}.
By itself, this is fine, as it is not possible
to mutate a \scode{List} object directly,
but as we will see, this
allows us to mutate the list indirectly.

\para{Covariance}
\begin{quote}
\begin{lstlisting}[numbers=none]
val covariance: List<B> = upcast\end{lstlisting}
\end{quote}
Next, \scode{getA} applies the covariance
of \scode{List} in order to upcast
\scode{List<A>} to \scode{List<B>}.
Since
any subclass of \scode{A} is also a subclass
of \scode{B}, 
a list of subclasses of \scode{A}
is also a list of subclasses of \scode{B}. 
This means that upcasting \scode{List<A>}
to \scode{List<B>} must be legal.
Indeed, the compiler does not issue
a warning or an error, and this line
runs successfully at run time due
to type erasure.
In fact, this line wouldn't crash
at run time
even if the list were not empty.

\begin{figure}[t]
\begin{lstlisting}[xleftmargin=1.5em, language=Kotlin,escapechar=@]
fun getA(): A {
    val list = mutableListOf<A>()
    val upcast: List<A> = list
    val downcast: MutableList<B> = 
                     upcast as MutableList@\label{line:downcastuncheckedcast}@
    downcast.add(B())
    return list[0]
}
\end{lstlisting}
\caption{A code snippet on which Kotlin's compiler issues an unchecked cast warning}
\Description{A code snippet on which Kotlin's compiler issues an unchecked cast warning}
\label{fig:uncheckedcast}
\end{figure}

\para{Downcast}
\begin{quote}%
\begin{lstlisting}[numbers=none]
val downcast: MutableList<B> = 
   covariance as MutableList\end{lstlisting}
\end{quote}
Now, \scode{getA} downcasts the resulting
object from \scode{List<B>} to \scode{MutableList<B>}.
This cast should definitely be legal statically,
as \scode{List} is a subclass of \scode{MutableList}.
Furthermore, 
as discussed in \Cref{sec:preliminaries},
at runtime, Kotlin exhibits type erasure, and so,
the only component of this cast that can be checked dynamically
is the class itself, which in this case is \scode{MutableList}.
This is sometimes referred to the cast being \emph{unchecked}.
Kotlin's compiler tries to reason 
about unchecked casts that could lead
to unsoundness, and only issues a warning 
if the type parameters of the
static type and the cast type are
different.
However, in our case, we're trying to cast a \scode{List<B>}
to \scode{MutableList<B>}, which means that the compiler
does not recognize this as an unchecked cast
and does not even issue a warning.
Contrast the code in \Cref{fig:mainexample}
with the code in \Cref{fig:uncheckedcast}:
both snippets compile and crash at the same place, 
but in \Cref{fig:uncheckedcast},
the compiler issues a warning due to 
the type parameter mismatch
in the cast on line \ref{line:downcastuncheckedcast}.

Ultimately, since the downcasted list object 
possesses the runtime type of \scode{MutableList},
the compiler does not issue
a warning or an error, and this line
runs successfully at runtime.
That is, in this case, the dynamic
type matches the cast type, and the static
type parameter matches the cast type's static
type parameter. 
This means that there's not really
a reasonable way to prevent this downcast
from being legal from a language design perspective.
However, the missing compiler warning stems
from a wrong assumption: unchecked casts
can only occur when the type parameters mismatch.
This matter is further discussed in
\Cref{sec:variations}.

\para{List mutation}
\begin{quote}%
\begin{lstlisting}[numbers=none]
downcast.add(B())
return list[0]\end{lstlisting}
\end{quote}
The next step is adding a \scode{B}
instance to the downcast list,
which has type \scode{MutableList<B>},
but crucially points to the same 
memory location as the original list,
which has type \scode{MutableList<A>}.
Since objects in the JVM are passed solely
by reference,
there's no memory issue
preventing us from adding a \scode{B}
object to a list of \scode{A}
objects, and this is purely a type checker
issue.
We can now extract the \scode{A}
object from the original list,
despite \scode{A} being
uninhabited.

\para{Post-mortem}
We have shown how Kotlin's
type system allows us to acquire
an instance of the uninhabited
type \scode{A} through
a series of coercions.
This must mean
that the type system is unsound.
Importantly, none of the steps shown
here will necessarily crash
the program at run time,
as none of these steps cause
a checked cast.
\Cref{subsec:checkcast}
details what operations
can be performed on the 
illegal object without
it being discovered.

\section{Discussion}\label{sec:variations}
\subsection{Who's to blame?}\label{subsec:blame}
When considering how this issue
can be fixed on a language design level,
we must consider the role
of every language feature
involved in this issue.
Note that unsoundness
is not caused by any
one of these language
features alone, but rather
by their interaction 
in ways that language developers
might not have expected.
This discussion may be also useful
to programmers wishing to avert
this issue in their own code.

\begin{figure}[t]
\begin{lstlisting}[xleftmargin=1.5em, escapechar=\#, style=sharpc]
class B {}
class A : B {}

interface List<out T>  
	{ /* ... */ }
interface MutableList<T> 
	: List<T>  { /* ... */ }
class ArrayList<T> 
	: MutableList<T> { /* ... */ }

A getA() {
    MutableList<A> list = new ArrayList<A>();
    List<A> upcast = list;
    List<B> covariance = upcast;
    MutableList<B> downcast = (MutableList<B>) covariance;#\label{line:csharpcast}#
    downcast.add(new B());
    return list.get(0);
}
\end{lstlisting}
\caption{A translation of \Cref{fig:mainexample} to C\#}
\Description{A translation of \Cref{fig:mainexample} to C\#}
\label{fig:csharp}
\end{figure}
\para{Type erasure}
The most obvious culprit is type erasure.
Languages without type erasure, like C\#, 
suffer from this problem to a lesser degree.
For example,
\Cref{fig:csharp} is a snippet of C\# code
that is analougous to \Cref{fig:mainexample}.
Since C\# does not have type erasure, 
the cast on line \ref{line:csharpcast}
of \Cref{fig:mainexample} would be a fully checked cast
and would fail immediately at run time
instead of returning a object in an invalid state.
However, the problem still exists, and \texttt{csc}
version 3.9.0-6.21124.20 still compiles the snippet
in \Cref{fig:csharp} without warnings.
This might mean that the problem is less serious
from a software engineering perspective
because it would be discovered earlier at run time,
but it still exists and causes C\#'s type system
to also be unsound.
Therefore, type erasure is \emph{not}
the only cause of unsoundness.

\begin{figure}[t]
\begin{lstlisting}[xleftmargin=1.5em, escapechar=\#]
open class B2<out T : Any> {
    lateinit var x: ^@UnsafeVariance^ T#\label{line:unsafevariance}#
        private set
}
\end{lstlisting}
\caption{A program that receives a compiler warning without the \scode{@UnsafeVariance} annotation on line \ref{line:unsafevariance}}
\Description{A program that receives a compiler warning without the \scode{@UnsafeVariance} annotation on line \ref{line:unsafevariance}}
\label{fig:unsafevariance}
\end{figure}

\begin{table*}[h!t]
\begin{tabular}{l@{\hspace{2em}}lc}
 Description & Example & \scode{CHECKCAST} placed?\\ \hline
 Expression statement & \kcode{bad} & \faRemove \\
 Explicitly-typed declaration & \kcode{val x: MyClass = bad} & \faCheck \\
 Implicitly-typed declaration & \kcode{val x = bad} & \faCheck \\
 Instance check & \kcode{bad is MyClass} & \faRemove \\
 Function call that accepts Any? & \kcode{println(bad)} & \faRemove \\
 Function call that accepts MyClass  & \kcode{(\{x: MyClass->0\})(bad)} & \faCheck \\
 Method call (or field access) on MyClass & \kcode{bad.childMethod()} & \faCheck \\
Method call (or field access) from a parent class & \kcode{bad.parentMethod()} & \faCheck \\
Return statement expecting return type Any? & \kcode{return bad} & \faRemove \\
Return statement expecting return type MyClass & \kcode{return bad} & \faCheck 	
\end{tabular}
\caption{Different cases in which the JVM instruction 
\scode{CHECKCAST MyClass} is (\faCheck) and is not (\faRemove)
inserted on \scode{bad}}
\Description{Different cases in which the JVM instruction 
\scode{CHECKCAST MyClass} is (\faCheck) and is not (\faRemove)
inserted on \scode{bad}}
\label{tab:checkcast}
\end{table*}

\para{Declaration-site variance}
This proof is fundementally linked
to declaration-site variance.
Declaration-site variance allows,
in a literal sense,
to ``hide''
non-conforming conversions.
These conversions are ``hidden''
from the compiler, which allows them
since they are allowed
by the language specification,
since the alternative is to forbid
\scode{MutableList} from inheriting
from \scode{List}.
Even worse,
they are hidden from the user,
which may not be aware at all
that this issue can occur,
since client code (like \Cref{fig:mainexample})
does not mention variance at all.
This is unlike use-site variance,
in which illegal conversions like
\scode{List<? extends T>}
to \scode{MutableList<T>}
would both be caught by the compiler
(while still allowing \scode{MutableList}
to inherit from \scode{List}),
and would be clearly visible in client code.

\para{Collections hierarchy}
Unlike Amin and Tate's proof,
our proof uses basic language features
and common conversions used in every-day code.
The real-world implications of this result
can be partially blamed on the design of
Kotlin's standard collections library, 
which are widely used.
We claim that this is a real problem
partially because
the Kotlin collections hierarchy
was not designed correctly:
the Liskov Substitution Principle~\citep{liskov}
should forbid the \scode{MutableList}
interface from extending
the \scode{List} interface.
This matter is further discussed in a blog
post by a Google developer~\citep{twomutables}.

\para{Implicit casts}
While implicit casts 
do not cause unsoundness in this case,
they help hide it from programmers:
as discussed in the next subsection,
the compiler fails to recognize
unchecked casts if the origin type
has the same type parameter as
the destination type.
Since type parameters can be altered implicitly
by covariance,
implicit casts worsen the problem
from the programmer's side.

\subsection{@UnsafeVariance}
Kotlin's developers were aware
that strict abidance to sound use of variance 
may lead to awkward code in real-world contexts
and introduced \scode{@UnsafeVariance}.
The \scode{@UnsafeVariance} annotation can be used
in cases where the compiler detects a use site of a type parameter
that contradicts its denoted variance \emph{in that same class},
and overrides this check, allowing the code to compile.
For example, the snippet in \Cref{fig:unsafevariance}
would not compile without the \scode{@UnsafeVariance}
annotation on line \ref{line:unsafevariance}.
This annotation represents the developer's acknowledgement
of potential unsoundness that may occur.
However, the annotation is \emph{not}
required if the unsound usage stems from non-variant
inheritance from a variant class.
For example, \scode{MutableList<E>\#add}
does not require the annotation, because
\scode{MutableList<E>} is not itself variant,
and despite \scode{List<E>} being variant.
One simple language design step that can be taken
to remedy the issue described in this paper
is requiring \scode{@UnsafeVariance}
to be placed on a non-variant
class inheriting from a variant class,
like so:
\begin{lstlisting}[numbers=none]
open class Base<out T>
class Derived<T> : @UnsafeVariance Base<T>()\end{lstlisting}

In fact, even if the implementation
of \scode{List} would have inherited
it directly instead of \scode{MutableList},
this issue would still exist: 
Kotlin collections are not implemented in Kotlin
but rather are part of the Java standard library,
on which Kotlin's rules are not enforced,
including the rule requiring \scode{@UnsafeVariance}.

\subsection{Compiler warnings}
The latest version of \texttt{kotlinc}
does not issue a warning
or error on this snippet,
neither does the Kotlin plugin for 
IntelliJ IDEA~\citep{intellij}.
\texttt{kotlinc} warns about
unchecked casts only if the result
type differs from the origin
type in its type parameter\footnote{The full implementation can be found here: 
\url{https://github.com/JetBrains/kotlin/blob/4c0ec0fae4fec75c793c5cf2aa4727bae8abe553/compiler/fir/checkers/src/org/jetbrains/kotlin/fir/analysis/checkers/FirCastDiagnosticsHelpers.kt\#L22}}.
For example, \texttt{kotlinc}
issues an unchecked cast warning
in line \ref{line:downcastuncheckedcast} of \Cref{fig:uncheckedcast},
which skips the covariance step.
This means that an \emph{implicit cast}
can make the difference
between at least having
a warning and its absence.

\para{Compiler-level patch}
The compiler can be patched
to partially fix this issue
by amending the erasure analysis
to use local data-flow analysis 
that checks which other types the casted
value was implicitly casted from.
For example, this DFA
would run when processing line \ref{line:downcast}
of \Cref{fig:mainexample} and
would return \scode{\{MutableList<A>, List<A>, List<B>\}}.

\noindent
The compiler can now use this information
to check if a cast from \emph{any}
of these would be unchecked.
This would only partially fix this issue, 
since these conversions may be split up
over several functions,
in which case a local DFA would not find
the other potential types.
In addition, this would be significantly
slower than the current implementation,
and would potentially compromise
interactivity in IDEs for large functions.

\subsection{CHECKCAST instruction}\label{subsec:checkcast}
Since class references
are stored as pointers, the compiler 
can perform sanity checks on
the dynamic type at any time.
This is done using the JVM
\scode{CHECKCAST} instruction,
as discussed in \Cref{sec:preliminaries}.
In cases where an unsoundness
is introduced in real-world
programs, we want to discover it
early. 
Unfortunately, the Kotlin
compiler does not always insert
the \scode{CHECKCAST}
instruction immediately
after a class instance is obtained.
\Cref{tab:checkcast} demonstrates
some of the cases in which
the \scode{CHECKCAST}
is (\faCheck) and is not (\faRemove)
inserted. 
If \scode{CHECKCAST} is not inserted
by the compiler,
the unsoundness will remain undiscovered for longer.
Furthermore, as seen in \Cref{tab:checkcast},
even instance checks will not crash,
though they will return the correct answer
according to their real type, i.e.,
the type whose constructor was called
to create that specific value.
Worse still, the compiler will insist
that the instance check is redundant.
That is, while adding \scode{println(list[0] is A)}
after line \ref{line:mutate} of \Cref{fig:mainexample}
would print \scode{false}, the compiler issues
a \scode{check for instance is always 'true'}
warning.

This may be surprising;
type errors should be 
discovered as near 
as possible to
where they were caused, 
and
developers may
expect that compiler warnings
would not contain wrong information.
Patching the Kotlin compiler to add
\scode{CHECKCAST} calls
at every first use or acquisition
of a value would cause
this issue to be discovered
earlier.

\subsection{Previously Known Unsoundness}\label{subsec:known}
Issue number KT-7972 in the Kotlin bug tracker\footnote{\url{https://youtrack.jetbrains.com/issue/KT-7972}}
describes a different problem, 
also related to variance and soundness. 
The relevant snippet is shown in \Cref{fig:kt7972}. 
This issue is purely related to variance and
type erasure.
Unsoundness here is caused because on line \ref{line:7972addanything},
Kotlin tries to infer a valid type for the \scode{E}
type parameter: the lowest
common ancestor of \scode{Int} and \scode{String}
which is \scode{kotlin.Any}.
Due to variance, \scode{list}
can be cast to \scode{List<Any>}
allowing the function call to be legal.
Next, due to type erasure, it is not possible
to check if the list is actually \scode{MutableList<E>}
for \scode{E = Any}, 
and so the instance checks succeeds.
Due to Kotlin's smart casts, inside the conditional
block, \scode{this} is automatically
casted to \scode{MutableList<Any>}.

This is a significantly less fundemental issue:
as mentioned in the bug tracker itself,
this can be solved by not allowing
generic (smart) casts from a variant
class to a non-variant class.
Note that this issue \emph{does not}
impact non-generic smart casts (like
\scode{this is MutableList<Int>})
as the Kotlin compiler already issues
an error when encountering them.
In addition, such an issue would not occur
in languages without type erasure,
as the instance check in line \ref{line:7972check}
would return \scode{false} as expected.

\begin{figure}[t]
\begin{lstlisting}[xleftmargin=0.7em, escapechar=\#]
private fun <E> List<E>.addAnything(element: E) {
    if (this is MutableList<E>) {#\label{line:7972check}#
        this.add(element)
    }
}

val list = mutableListOf<Int>()
list.addAnything("string")#\label{line:7972addanything}#
val bad = list[0]
\end{lstlisting}
\caption{Code demonstrating issue number KT-7972 in the Kotlin bug tracker, demonstrating how variance, type erasure 
and smart casts violate soundness.}
\Description{Code demonstrating issue number KT-7972 in the Kotlin bug tracker, demonstrating how variance, type erasure 
and smart casts violate soundness.}
\label{fig:kt7972}
\end{figure}

\section{Conclusion}
Type systems are expected
to protect developers from
performing illegal operations
on values.
When a type system is unsound,
it cannot guarantee that
well-typed programs cannot
cause run time type errors,
which can lead to unexpected
crashes.

This paper described a
counterexample to soundness of the Kotlin type system
based on a novel combination of language
features. 
Our counterexample made use
of type erasure,
declaration-site variance
and Kotlin's
variance inheritance rules;
unlike previously known examples,
our counterexample is fundemental
and cannot be fixed without major changes
to the language.

Finally, we discussed
how different aspects
of Kotlin affect this issue;
on when it would be discovered
at runtime; 
and a possible
fix to the compiler
that can help detect
this source of unsoundness
in some cases.

\bibliographystyle{acm}

\bibliography{refs}

\end{document}